\documentclass{webofc}
\usepackage[varg]{txfonts}   
\begin{document}
\title{Simulation of Collective Excitations in the Stack\\ of Long Josephson Junctions}
%
%

\author{\firstname{Ilhom} \lastname{Rahmonov}\inst{1,2}\fnsep\thanks{\email{rahmonov@theor.jinr.ru}} \and
        \firstname{Yuri} \lastname{Shukrinov}\inst{1,3}\fnsep\thanks{\email{shukrinv@theor.jinr.ru}}\and
        \firstname{Pavlina} \lastname{Atanasova}\inst{4}\and
        \firstname{Elena} \lastname{Zemlyanaya}\inst{3,5}\and
        \firstname{Oksana} \lastname{Streltsova}\inst{5,6}\and
        \firstname{Maxim} \lastname{Zuev}\inst{5}\and
        \firstname{Andrej} \lastname{Plecenik}\inst{7}\and
        \firstname{Akinobu} \lastname{Irie}\inst{8}
}

\institute{BLTP, Joint Institute for Nuclear Research, Dubna 141980, Russia
\and
           Umarov Physical Technical Institute, Tajik Academy of Science, Dushanbe 734063, Tajikistan
\and
           Dubna State University, Dubna 141980, Russia
\and
           Plovdiv University of ``Paisi Khilendarsky'', 4003, Plovdiv, Bulgaria
\and
           LIT, Joint Institute for Nuclear Research, Dubna 141980, Russia
           \and
Peoples' Friendship University of Russia (RUDN University), Moscow,
Russia
\and
           Department of Experimental Physics, Comenius University, Bratislava, Slovakia
\and
           Utsunomiya University, Utsunomiya 321-8585, Japan
           }

\abstract{
  The phase dynamics of the stack of long Josephson junctions has been studied. Both inductive and capacitive couplings between Josephson junctions
have been taken into account in the calculations. The IV--curve, bias current dependence of radiation power and dynamics of each JJs of the stack have been investigated. The coexistence of the charge traveling wave and fluxon states has been observed. This state can be considered as a new collective excitation in the system of coupled Josephson junctions. We demonstrate that the observed collective excitation leads to the decreasing of radiation power from the system.
}

\maketitle
\section{Introduction}
\label{intro}
The layered high-Tc superconducting materials such as Bi$_{2}$Sr$_{2}$CaCu$_{2}$O$_{8+\delta}$  (BSCCO) can be considered as a stack of coupled Josephson junctions (JJs)~\cite{kleiner92}. The interest to the investigation of this system is caused by its rich nonlinear properties and different applications, particularly this system is one of the promising object of superconducting electronics~\cite{yurgens00}. The JJs stack demonstrates a series of interesting properties such as parametric resonance~\cite{shukrinov07-prl,shukrinov08-prb,rahmonov-jetpl14-eng}, chaotic features~\cite{shukrinov14-chaos} and in this system the fluxons~\cite{fulton72,pedersen84,mclauhlin78} and collective excitations~\cite{kleiner00,matsuda95} can arise. Coherent terahertz electromagnetic radiation from this system provides wide possibilities for various applications~\cite{ozyuzer07,welp13}. The effect of collective excitations on radiation power from the stack is not studied in detailed yet. In this paper, we investigate the IV-curve of stacked JJs  and radiation power from the stack taking into account the inductive and capacitive couplings~\cite{sakai93, machida04-sakai} and the diffusion current~\cite{shukrinov-PhysC06-mahfouzi,shukrinov-jetp12-eng,rahmonov-jetpl14-eng}.

The system of equations which describes the phase dynamics of the coupled long JJs stack in the normalized quantities can be written as following \cite{sakai93,machida04-sakai,rahmonov-jetp17}

\begin{equation}
\label{sys_eq1}
\left\{\begin{array}{ll}
\displaystyle\frac{\partial\varphi_{l}}{\partial t}=D_{c}V_{l}+s_{c}V_{l+1}+s_{c}V_{l-1},
\vspace{0.2 cm}\\
\displaystyle
\frac{\partial V_{l}}{\partial t}=\sum\limits^{N}_{k=1}\pounds_{lk}^{-1}\frac{\partial^{2}\varphi_{k}}{\partial x^{2}}-\sin\varphi_{l}+ \beta \frac{\partial\varphi_{l}}{\partial t}+I,
\end{array}\right.
\end{equation}

\noindent where $\varphi_{l}$ and $V_{l}$ is phase difference and voltage of $l$th JJ, respectively. $I$ is the bias current normalized to the critical current $j_{c}$ and $\beta$ is a dissipation parameter. Here $D_{c}=1+ (2\lambda_{e}/d_{I}) \coth(d_{s}/\lambda_{e})$ is the effective electrical thickness of JJ normalized to the insulating layer thickness $d_{I}$, $s_{c}=-\lambda_{e}/[d_{I}\sinh(d_{s}/\lambda_{e})]$ is the capacitive coupling parameter, $d_{s}$ is the thickness of superconducting layer, and $\lambda_{e}$ is Debye screening length. In this system of equations the voltage is normalized to $V_{0}=\hbar\omega_{p}/(2e)$, where $\omega_{p}=\sqrt{8\pi d_{I}ej_{c}/(\hbar \varepsilon)}$ is the plasma frequency of JJ, and $\varepsilon$ is the dielectric constant of the insulating layer. The time $t$ and coordinate $x$ are normalized to plasma frequency $\omega_{p}^{-1}$ and Josephson penetration depth $\lambda_{J}$, respectively. The matrix of inductive coupling $\pounds$ has the form
\[
\hat{\pounds}=\left(\begin{array}{ccccccc}
 1   & S  & 0  & \cdots &     &     &     S\\
     &    &    &     & \cdots &     &    \\
 \cdots & 0  & S  & 1   & S   & 0   & \cdots \\
     &    &    &     &     &     &    \\
 S   &    &    &     & 0   & S   & 1 \\
\end{array}\right),
\]
where $S=s_{\pounds}/D_{\pounds}$ is the inductive coupling parameter,
$s_{\pounds}=-\lambda_{L}/\sinh(d_{s}/\lambda_{L})$, $D_{\pounds}=d_{I}+2\lambda_{L}\coth(d_{s}/\lambda_{L})$ is the effective magnetic thickness of JJ, and $\lambda_{L}$ is the London penetration depth. The valid values of the inductive coupling parameter $S$ are in the range $S \in (-0.5, 0)$. The initial conditions for the system of equations (\ref{sys_eq1}) are $\varphi_{l}(x,0)=0$ and $V_{l}(x,0)=0$. The boundary conditions in the $x$ dirrection given by the external magnetic field $\partial\varphi_{l}/\partial x|_{x=0,L}=B_{ext}$. Here magnetic field is normalized to $B_{0}=\hbar c/2eD_{\pounds}$. In the $z$ dirrection we use the periodic boundary condition: in the case $l=N$, $\varphi_{l+1}=\varphi_{1}$, $V_{l+1}=V_{1}$; in the case $l=1$, $\varphi_{l-1}=\varphi_{N}$, $V_{l-1}=V_{N}$.

Simulations are based on a numerical solution of a system of nonlinear partial differential equations by the fourth order Runge--Kutta method, a finite-difference approximation, and the MPI technique for parallel implementation. The details of simulation of IV-curve and other characteristics are discussed in a number of our previous papers~\cite{shukrinov-lncs12, rahmonov-jetpl14-eng, rahmonov-jetp17}. Power of radiation from the stack is simulated using the expression $P=V^{2}_{ac}/R_{Z}$~\cite{krasnov,rahmonov-jetp17}, where $V_{ac}=d_{I}E_{ac}$, $E_{ac}$ is AC part of electric field, $R_{Z}=(d_{I}/W)Z$, $W$ is width of JJ, $Z$ is impedance of radiation. The details of simulation of radiation power are discussed in Refs.\cite{krasnov, rahmonov-jetp17}. Magnetic field in the JJs is calculated using the expression $B_{l}= B_{0}\sum^{N}_{k=1}\pounds_{lk}^{-1}\partial\varphi_{k}/ \partial x$. The charge in superconducting layers is calculated using the expression $Q_{l}(x,t)=Q_{0}[V_{l}(x,t)-V_{l-1}(x,t)]$~\cite{rahmonov-jetpl14-eng}, where
$Q_{0}=\varepsilon V_{0}/ 4\pi d_{s} d_{I}$.

\section{Result and discussions}
\label{sec-2}
\begin{figure}[h!]
\centering
\sidecaption
\includegraphics[width=6.5cm,clip]{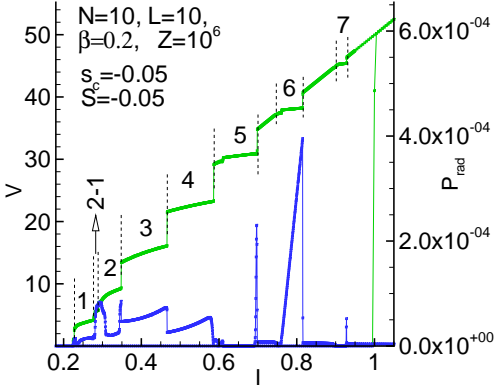}
\caption{IV-curve together with bias current dependence of radiation power calculated for the stack of $N=10$ JJs with model parameters $D_{c}=1.1$, $s_{c}=-0.05$, $S=-0.05$ and $\beta$.
Vertical dashed lines show the boundaries of ZFS. For each ZFS the number of fluxons is indicated.}
\label{cvc}       

\vspace{-1\baselineskip}
\end{figure}

In Fig.\ref{cvc} the IV-curve and power of radiation as a function of bias current for the stack of ten JJs are presented. Calculation is provided for the stack with disspation parameter $\beta=0.2$, inductive coupling $S=-0.05$, capacitive coupling $s_{c}=-0.05$ and effective electric thickness $D_{c}=1.1$. IV-curve demonstrates seven zero field steps (ZFS)~\cite{fulton72,pedersen84,mclauhlin78}, which are formed due to the appearance of fluxons (kinks and antikinks). The vertical dashed lines show the boundary of regions of IV-curve where fluxons appear. For each region the corresponding number of fluxons is indicated. Between the regions of IV--curve with one and two fluxons near some ZFSs in compare with a case of single JJ appears an additional branch due to the different number of fluxons in the each JJs of the stack. This situation was discussed in Ref.~\cite{rahmonov-jetp17}. In the case of single JJ a significant radiation power is observed in the regions of IV--curve corresponding to the ZFS \cite{rahmonov-jetp17}. In the case of the stacked JJs the radiation power is closed to the zero near some ZFSs in compare with a case of single JJ (in our case ZFS with 5 and 7 fluxons).

\begin{figure}[h!]
\centering
\includegraphics[width=6.5cm,clip]{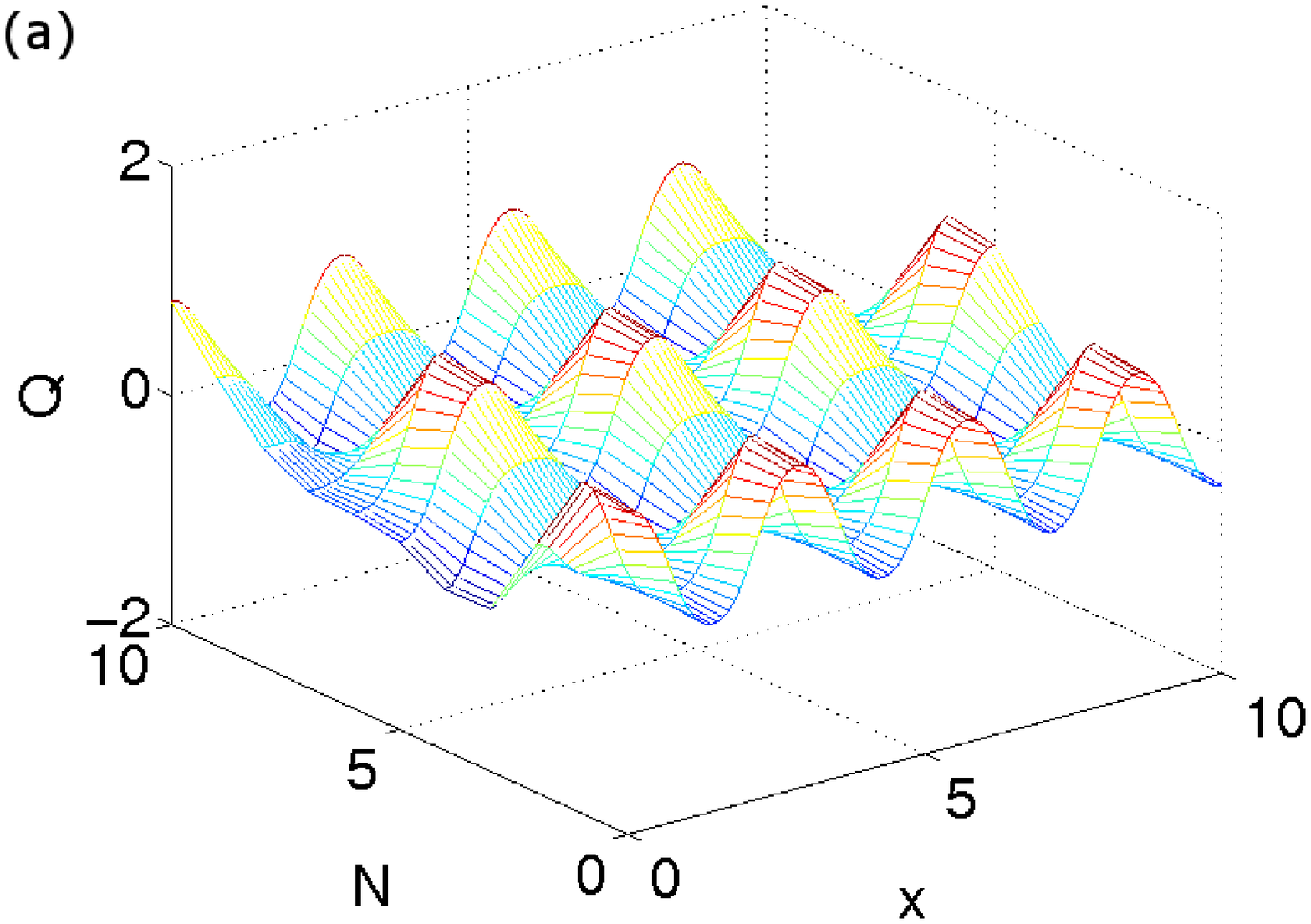}
\hspace{1cm}
\includegraphics[width=5cm,clip]{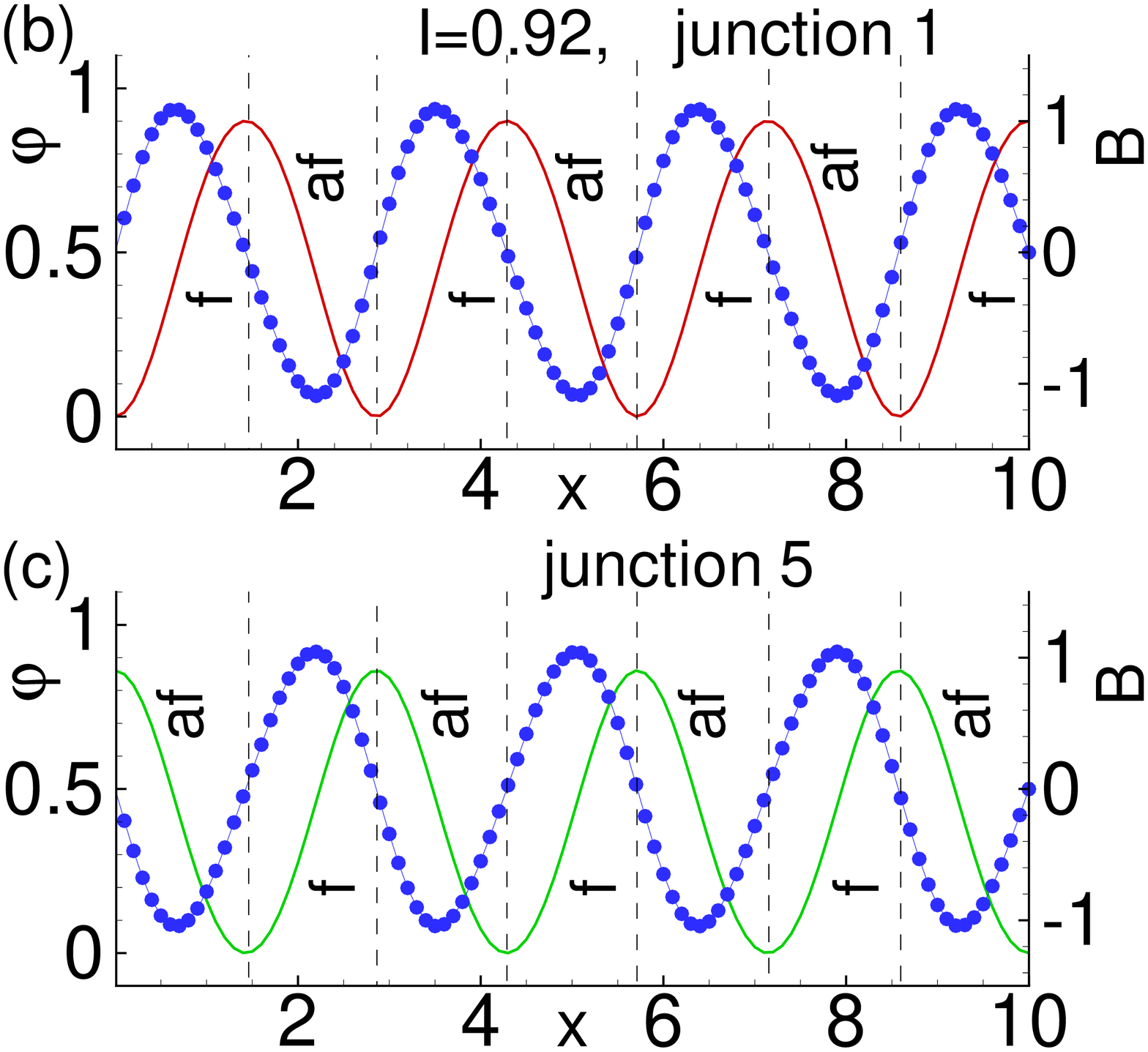}
\caption{(a) Charge distribution in superconducting layers along the coordinate and stack of JJs at $I=0.92$ in fixed time moment; (b) Distribution of
 phase difference (solid line) and magnetic field (doted line) along the coordinate at $I=0.92$ in fixed time moment for first JJ. The boundaries of fluxons and antifluxons are shown with the vertical dashed line; (c) The same as in the case (b) for fifth JJ.}
\label{charge}       

\vspace{-1\baselineskip}
\end{figure}

In order to explain such behavior of stacked JJs we  have investigated the dynamics of each JJs of stack and calculate charge dynamics in superconducting layers in the above mentioned region of IV--curve. Figure~\ref{charge}(a) demonstrates the distribution of charge in superconducting layers along the stack and coordinate, for fixed time moment at $I=0.92$. It can be seen that along the stack wave--like behavior of charge  is realized. Along the coordinate we see the fluxons. The analysis of dynamics of each JJs of the stack demonstrates that along the stack a charge traveling wave (CTW) is appeared, i.e. the CTW and fluxons are coexisted. Such dynamical state can be considered as a new collective excitation in the stacked JJs.

A reasonable question is appeared here: why the observed collective excitation leads to the decreasing of radiation power? In order to explain this phenomena we have analyzed dynamics of phase difference and magnetic field for each JJs of the stack. In Fig.~\ref{charge}(b) a spatial distribution of the phase difference (solid line) and magnetic field (doted line) at the fixed time moment in the first JJ at the current value $I=0.92$ are shown, which corresponds to the seventh ZFS. This dependence demonstrates seven fluxons (four fluxons and three antifluxons). The same characteristics for the middle of stack, i.e., in the fifth JJ are presented in Fig.~\ref{charge}(c). In compare with the first JJ (see Fig.~\ref{charge}(b)), here we can see four antifluxons and three fluxons, i.e., the opposite situation. This may us conclude that first and fifth JJs are in opposite phase. This circumstance results  exactly to the zero value of averaged radiation power.


%
%
%

\section{Conclusions}
\label{sec-3}

In this paper we have investigated the structure of IV-curve of JJs and radiation power from the stack of JJs. We demonstrate that in the stack of JJs a charge traveling wave can appear in the ZFS, i.e. fluxons coexist with the charge traveling wave. This indicates the appearance of a new collective excitation in the system of coupled JJs. We note that such collective excitation on in stack of JJs has a significant influence on radiation power from the stack of JJs. Particularly, it leads to the zero radiation power from the stack. We assume that the obtained results can be used in analysis of experimental IV--curve. 

\vspace{-0.5\baselineskip}
\section*{Acknowledgement}

The study was funded by the RFBR, according to the research project 15--29--01217, Heisenberg--Landau programm and JINR-Slovakia collaboration, and by the Ministry of Education and
Science of the Russian Federation (the Agreement number 02.a03.21.0008).

\vspace{-0.5\baselineskip}

{}

\end{document}